\documentclass[%
 aip,a4paper,
 amsmath,amssymb,
reprint
]{revtex4-1}

\usepackage{graphicx}% Include figure files
\usepackage{dcolumn}% Align table columns on decimal point
\usepackage{bm}% bold math
\usepackage[T1]{fontenc}
\usepackage{mathptmx}

\renewcommand{\[}{\begin{equation}}
\renewcommand{\]}{\end{equation}}
\def\bea{\begin{eqnarray}}
\def\eea{\end{eqnarray}}
\def\nn{\nonumber\\}
\newcommand{\equ}[1]{Eq.~(\ref{#1})}
\newcommand{\eqs}[2]{Eqs.~(\ref{#1}) and (\ref{#2})}

\def\R{{\bf R}}

\def\zz{{\mathfrak z}}

\newcommand{\ei}[1]{{\rm e}^{i #1}}
\newcommand{\emi}[1]{{\rm e}^{-i #1}}

\def\ket#1{\vert#1\rangle}
\def\ev#1{\langle#1\rangle}
\def\me#1#2#3{\langle#1| \, #2 \, |#3\rangle}
\begin{document}
\preprint{AIP/123-QED}

\title{Faraday law, oxidation numbers, and ionic conductivity: \\ The role of topology}
% Force line breaks with \\

\author{Raffaele Resta}
% \altaffiliation[Also at ]{Donostia International Physics Center, 20018 San Sebasti{\'a}n, Spain}%!TEX encoding = UTF-8 UnicodeLines break automatically or can be forced with \\
%\author{B. Author}%
\email{resta@iom.cnr.it}
\affiliation{ 
Istituto Officina dei Materiali IOM-CNR, Strada Costiera 11, 34151 Trieste, Italy%\\This line break forced with \textbackslash\textbackslash
}%

\homepage{http://www-dft.ts.infn.it/\~{}resta/.}
\affiliation{Donostia International Physics Center, 20018 San Sebasti{\'a}n, Spain%\\This line break forced% with \\
}%

\date{\today}% It is always \today, today,

\begin{abstract}
Faraday's experiment measures---within a modern view---the charge adiabatically transported over a macroscopic distance by a given nuclear species in insulating liquids: the reason why it is integer is deeply rooted in topology. Whole numbers enter chemistry in a different form: atomic oxidation states. They are not directly measurable, and are determined instead from an agreed set of rules. Insulating liquids are a remarkable exception: Faraday's experiment indeed measures the oxidation numbers of each dissociated component in the liquid phase, whose topological values are unambiguous. Ionic conductivity in insulating liquids is expressed in terms of the autocorrelation function of the fluctuating charge current at a given temperature in zero electric field; topology plays a major role in this important observable as well. The existing literature deals with the above issues by adopting the independent-electron framework; here I provide the many-body generalization of all the above findings, which furthermore allows for compact and very transparent notations and formulas. 
\end{abstract}

\maketitle

\section{Introduction}

Faraday's first law of electrolysis addresses insulating liquids in electrolytic cells, and can be recast in modern terms as follows. When a macroscopic number $N$ of nuclei of a given chemical species passes from one electrode to the other, the transported electrical charge is an integer multiple of $N$ times $e$. The law is additive: it concerns each different chemical species in the cell independently. It addresses ionic conduction in electrolytes, including molten salts; it does not apply to metallic conduction, which is unrelated to the motion of nuclei.

Faraday's discovery dates from the 1830s, decades before the existence of atoms became accepted. Remarkably, he wrote:\cite{Faraday} ``Although we know nothing of what an atom is, yet we cannot resist forming some idea of a small particle...''. What he actually measured were charge-to-mass ratios; when scaled to the actual atomic masses, Faraday's ``equivalent numbers'' can be regarded as the archetypical definition of oxidation numbers (also known as oxidation states) in insulating liquids. I also stress that such definition is not static: Faraday's numbers are---within a modern quantum-mechanical view---dynamical properties in the adiabatic limit.

In modern chemistry, the oxidation number of an atom in a molecule or in a solid is an integer determined by an agreed set of rules,\cite{Walsh18} and is often related in a postdictive\cite{rap142}---{\it not} predictive---way to some measurable properties, thus facilitating the interpretation of several experimental observations.\cite{Walsh18,rap142,rap_a29,Sit11,Jiang12} Nowadays we know ``what an atom is'', but we also know that solids and liquids are not assemblies if ions; they are assemblies of atoms, having ionic character only because neighboring atoms have different electronegativity.\cite{Appelbaum76} Literally dozens of electronegativity scales and/or definitions of atomic charges have been proposed in the literature,\cite{Meister94} none of them yielding integer values. 

The reason why integer oxidation numbers in electrolytes are measurable quantities is deeply rooted in topology and in a milestone 1983 paper by D. Thouless;\cite{Thouless83} 
while this paper only addresses solid-state issues, its relevance for understanding quantized charge transport in ionic liquids was emphasized shortly after in a very little quoted---and presumably also little known---paper by Pendry and Hodges.\cite{Pendry84} Since then, a few authors have built upon their result and endorsed the topological nature of oxidation numbers in solids and liquids.\cite{rap_a29,rap142,Jiang12,Grasselli19} The main message from topology is that the ionic charges behave as integer only when adiabatically transported in a  macroscopic system; no trace of quantization may appear in a ``snapshot'' of either a solid or a liquid at a given time. The quantized charge transport may occur either in a real experiment (as in Faraday's) or in a ``gedanken'' experiment, nowadays called a ``Thouless pump''.

An outstanding 2019 paper by Grasselli and Baroni\cite{Grasselli19} has extended the scope of topology in the context of charge transport in insulating liquids. Besides the amount of transported charge, another basic observable is the value of ionic conductivity at a given temperature, which is provided within classical statistical mechanics by the fluctuation-dissipation theorem, as expressed in the Green-Kubo formula, based on the autocorrelation function of the fluctuating charge current.\cite{Hansen,Frenkel} Owing to the adiabatic approximation, the nuclear motion is classical even within ab-initio molecular dynamics, where the quantum nature of the electrons is fully accounted for. The contribution to the current from the motion of each nucleus at a given time is quite different from the one of a scalar integer charge---the topological oxidation number times $e$--moving on the same trajectory. According to Ref. \onlinecite{Grasselli19}, topology warrants that, when the real current is replaced with the fictitious current carried by the said integer charges (along the ab-initio trajectories), the Green-Kubo formula yields the correct ionic conductivity.

Charge transport in condensed matter has a close relationship to the theory of polarization (not ``modern'' anymore), developed along the 1990s.\cite{rap73,King93,Vanderbilt93,Ortiz94,rap100} In fact the current density (due to electrons and nuclei) which flows in a macroscopic insulator while the nuclei are adiabatically displaced equals by definition the time-derivative of macroscopic polarization. The formal definition of oxidation numbers proposed in Ref. \onlinecite{Grasselli19} for insulating liquids is indeed rooted in polarization theory, in its independent-electron (i.e. density-functional) version.\cite{Vanderbilt}
At variance with the previous literature, here I thoroughly address charge transport in insulating liquids in a many-body framework, thus generalizing the main results of Refs. \onlinecite{Pendry84} and \onlinecite{Grasselli19} beyond the independent-electron level. The present many-body approach naturally yields a compact and very perspicuous formulation.

In Sec. \ref{sec:pump} I introduce the closed system---made of classical nuclei and quantum electrons---adopted throughout this work in order to represent an electrolytic cell and  charge transport therein. As an introduction to the following formalism, in Sec. \ref{sec:classical} I address classical transport, in order to show how its essential features are conveniently recast in terms of a phase angle. The quantum electronic transport is dealt with in Sec. \ref{sec:qt}; the theory is also based on a phase angle,\cite{rap100,rap162} now a property of the many-body wavefunction, which sometimes is dubbed ``single-point Berry phase''. In Sec. \ref{sec:topo} the main findings of Refs. \onlinecite{Pendry84,Grasselli19} are retrieved in the present compact many-body framework in order to arrive at an unambiguous definition of oxidation numbers in the insulating liquid. The special case of noninteracting electrons is dealt with in Sec. \ref{sec:indep}, which relies on the Wannier-function formulation of polarization theory.\cite{Vanderbilt} The recent Grasselli-Baroni major advance in conductivity theory is generalized in Sec. \ref{sec:cond}. In Sec. \ref{sec:more}  I discuss the topological oxidation numbers, as emerging from the present analysis of Faraday's experiment, in their relationship to their more conventional definition in the chemistry literature; a remarkable pitfall is discussed in the following Sec. \ref{sec:pit}. In Sec. \ref{sec:conclu} I present some concluding remarks. In Appendix \ref{sec:insu} I outline the main points of the theory of the insulating state as adopted here;\cite{rap107,rap_a25,Stella11,rap_a33} in Appendix \ref{sec:infra} I present a different physical property whose theory partly overlaps with the present theory: infrared absorption in insulating liquids. A few somewhat advanced mathematical considerations have been confined to Appendix \ref{sec:chern}.

\section{Electrolytic cell as a Thouless pump} \label{sec:pump}

A liquid is an assembly of nuclei and electrons; when a nucleus takes off from one electrode and lands on the other in the electrolytic cell it pumps some electronic charge across the cell. The electronic current, driven by the electromotive force, flows through the closed circuit which includes the battery; owing to continuity, we may  address transport by monitoring the passage of nuclei and electrons across an ideal section of the cell. 

We wish to describe the electrons in the liquid by means of a finite (and closed) quantum system: in order to account for dc steady-state currents it is then mandatory to adopt periodic boundary conditions (PBCs). We therefore assume that the liquid is contained in a macroscopic cylindrical vessel of section $A$ and length $L$ (along $x$). By adopting PBCs with period $L$,  the $x$ coordinate becomes equivalent to the angle $\varphi = 2\pi x/L$. The system comprises therefore a central cell and its replicas in infinite number, thus acquiring the topology of a torus. 

One cycle of the pump consists in adiabatic transporting one nucleus from a given position in the central cell to its replica position in the nearby cell, thus making one loop around the torus; the transported charge $Q$ is the bare nuclear charge plus the time-integrated electronic current across the vessel's section. It will be shown next that, because of topological reasons, an integer electronic charge is pumped: as said above, the process goes nowadays under the name of Thouless pump.

The Thouless pump was first formulated in 1983 at the independent-electron level.\cite{Thouless83} The present work is deeply rooted in its many-body generalization, presented by Niu and Thouless\cite{Niu84}  in 1984. A comprehensive review of related topics can be found in Ref. \onlinecite{Xiao10}.

\section{Classical transport as a winding number}  \label{sec:classical}

We start considering the simplest classical model of a binary molten salt: point-like ions of charges $\pm qe $, interacting e.g. via two-body forces (Coulomb and short-range).\cite{Hansen,Hansen75} Clearly Faraday law requires integer values of $q$, in which case it is trivially fulfilled by construction. If $\R_\ell^{(+)}(t)$ and $\R_\ell^{(-)}(t)$ are the instantaneous positions of the $N$ positive and $N$ negative ions, one could write the macroscopic electrical current density at time $t$, for each chemical species, as
\[ {\bf j}^{(+)}(t) = \frac{q e}{LA} \sum_{\ell=1}^N \dot{\R}_\ell^{(+)}(t), \quad  {\bf j}^{(-)}(t) = - \frac{qe}{LA} \sum_{\ell=1}^N \dot{\R}_\ell^{(-)}(t) . \label{trivial} \] 
In order to deal with the quantum-mechanical case addressed in the following,
it proves convenient instead to express the adiabatic electrical currents density as the time derivative of macroscopic polarization ($x$-component thereof): $j_x^{(\pm)}(t) = d P_x^{(\pm)}(t)/dt$. 

When written in terms of the angles $2\pi x^{(+)}_\ell(t)/L$,
the $x$ component of the cationic current is \bea j_x^{(+)}(t) &=& \frac{d P^{(+)}(t)}{dt} = \frac{e}{2 \pi A} \frac{d}{dt}  \mbox{Im ln } \zz^{(+)}(t) \nn  \zz^{(+)}(t) &=& \ei{\frac{2\pi q}{L} \sum_{\ell=1}^N X_\ell^{(+)}(t)} , \label{smart} \eea where $X_\ell^{(+)} \equiv R_{\ell,x}^{(+)}$, and analogously for the anionic current $j_x^{(-)}(t)$; \equ{smart} is clearly additive, given that \[ \mbox{ln } \ei{\frac{2\pi q}{L} \sum_{\ell=1}^N X_\ell^{(+)}(t)} = \sum_{\ell=1}^N   \mbox{ln } \ei{\frac{2\pi q}{L} X_\ell^{(+)}(t)} . \] \ We notice that according to \equ{smart} macroscopic polarization is a multivalued quantity, as indeed mandated by the theory of polarization.\cite{Vanderbilt93,Vanderbilt} 

The two complex numbers 
$\zz^{(\pm)}(t)$ trace a path on the unit circle in the complex plane; the branch choice in ``Im ln'' is made in such way that each current is a continuous function of $t$.
In terms of the phase angles $\gamma^{(\pm)}(t) = \mbox{Im ln } \zz^{(\pm)}(t)$, the charge transported across a section of the toroidal vessel in time $T$ by the nuclei of each chemical species is proportional to the accumulated angle: \[ Q^{(\pm)}(T) = A \int_0^T  dt \; j_x^{(\pm)}(t) = \frac{e}{2\pi} [\; \gamma^{(\pm)}(T) - \gamma^{(\pm)}(0) \;] .\] 
In the special case where the trajectories $\R^{(\pm)}_\ell(t)$ are periodic in time with period $T$, the net charges $Q^{(\pm)}(T)$ transported across the section by the ions of each species coincides  with the winding numbers of $\zz^{(\pm)}(t)$ in the complex plane times $e$.

Finally, the total classical current which flows through the vessel can be expressed in terms of the time-derivative of a phase angle: \bea I_x(t) &=& A j_x(t) = \frac{e}{2\pi}\dot \gamma(t) , \label{icl}  \\ \gamma(t) &=& \mbox{Im ln } \ei{\frac{2\pi q}{L} \sum_{\ell=1}^N [\, X_\ell^{(+)}(t) -  X_\ell^{(-)}(t) \,]} \label{gcl} . \eea

\section{Quantum transport} \label{sec:qt}

The $\ell$-th nucleus, when infinitesimally moved, displaces an effective charge due to the nucleus itself and to the polarization of the electronic distribution. The concept is exactly cast into the (instantaneous) Born effective charge tensor\cite{Baroni01,Vanderbilt} $eZ^*_{\ell,\alpha\beta}(t)$, where Greek indices indicate Cartesian coordinates. The Born charges go also under the name of infrared charges and (in quantum chemistry) of atomic polar tensors; the $\ell$-th tensor depends on the environment at time $t$, it is anisotropic, and is not even symmetric in general. Within the classical ionic model discussed above $Z^*_{\ell,\alpha\beta}(t)$ must be identified with $\pm q \delta_{\alpha\beta}$ at all times. If there are $N$ nuclei in the vessel, 
the macroscopic electrical current which flows through the torus is \[ I_x(t) = A \, j_x(t) = \frac{e}{L} \sum_{\ell=1}^N \sum_\beta  Z^*_{\ell,x\beta}(t) \dot R_{\ell,\beta}(t) ; \label{born} \] the same adiabatic current flows in average though a given section of the vessel.
The exact expression of \equ{born} is additive, but clearly the charge carried by each nucleus is neither scalar nor integer. Faraday's law will manifest itself only when a nucleus is carried over a macroscopic distance.

We start addressing the electronic term only.
Within the adiabatic approximation, the many-electron wavefunction obeys a time-dependent Schr\"odinger equation, whose one-body potential is determined by the nuclear trajectories. Let $\ket{\Psi(t)}$ be the instantaneous adiabatic ground eigenstate at time $t$ of a system of $N^{(\rm el)}$ electrons, within the same periodic boundary conditions as above: $\ket{\Psi(t)}$ is periodic with period $L$ in the $x$ coordinate of each electron, independently.
The adiabatic electronic current density at time $t$ is:\cite{rap100,rap_a20} \bea j_x^{(\rm el)}(t) &=& \frac{e}{2 \pi A} \frac{d}{dt}  \lim_{L\rightarrow\infty} \mbox{Im ln } \zz^{(\rm el)}(t) \nn  \zz^{(\rm el)}(t) &=& \me{\Psi(t)}{\emi{\frac{2\pi}{L} \sum_{j=1}^{N^{(\rm el)}} x_j}}{\Psi(t)} ; \label{rapix} \eea the matrix element is the expectation value of a unitary operator, ergo $\zz^{(\rm el)}(t)$ is a complex number whose modulus is no larger than one. The limit will be omitted in the following formulas, given that both $L$ and $A$ are macroscopic.

At this point we may add the contribution to the electrical current from the $N$ classical point-like nuclei: it is enough to adopt \equ{smart}, where the bare nuclear charges  $e Z_\ell$ replace the ionic charges $\pm q e$: \[  \zz(t) = \me{\Psi(t)}{\ei{\frac{2\pi}{L}(\sum_{\ell=1}^N Z_\ell X_\ell -\sum_{j=1}^{N^{(\rm el)}} x_j)}}{\Psi(t)} \label{zz} , \]
where, owing to charge neutrality, $N^{(\rm el)} = \sum_{\ell=1}^N Z_\ell$. 
Analogously to the classical case of \eqs{icl}{gcl}, the total current is expressed by means of the time-derivative of a phase angle:
\bea I_x(t) &=& \frac{e}{2 \pi} \dot\gamma(t) = \frac{e}{2 \pi} \frac{d}{dt} \mbox{Im ln } \zz(t) \label{rapix2} \\ \gamma(t) &=& \mbox{Im ln }\me{\Psi(t)}{\ei{\frac{2\pi}{L}(\sum_{\ell=1}^N Z_\ell X_\ell -\sum_{j=1}^{N^{(\rm el)}} x_j)}}{\Psi(t)} .  \label{single} \eea The phase $\gamma(t)$ (electronic term thereof) often goes in the literature under the somewhat oxymoronic name of ``single-point Berry phase''.\cite{rap_a20}

We stress that \eqs{rapix2}{single} are equivalent to \equ{born}, although in the $L \rightarrow \infty$ limit only; the equivalence proves that \equ{rapix2}, despite not being additive in-form, is additive indeed. While $\gamma(t)$ itself is {\it not} additive, its time-derivative is such, as a consequence of linear response: the total current is the sum of the currents induced by infinitesimally displacing each nucleus one at a time. This ``dynamical'' additivity can be exploited by defining \bea \dot \gamma_\ell(t) &=& \frac{2\pi}{L} \sum_\beta  Z^*_{\ell,x\beta}(t) \dot R_{\ell,\beta}(t) \nn \gamma_\ell(t) &=& \int_0^t dt' \; \dot \gamma_\ell(t') ,  \label{dyn} \eea such that \[ \gamma(T) - \gamma(0) = \sum_{\ell=1}^N \gamma_\ell(T) . \label{add} \] The total accumulated angle is therefore unambiguously decomposed into the sum of the angles accumulated by each nucleus; each $\gamma_\ell(t)$, nonetheless, depends on the history of the system via the trajectories $\R_\ell(t)$.

The total electrical charge which crosses a vessel's section in time $T$ is thus in average \[ Q(T) = \frac{e}{2\pi} [\gamma(T) - \gamma(0)]; \] 
as in the classical case, if the nuclear coordinates $\R_\ell(t)$ are periodic in time, the transported charge is integer, and is measured by the winding number of $\zz(t)$ in the complex plane times $e$. The $3N$-dimensional space of the $\R_\ell(t)$ coordinates is a torus (in the $x$-coordinate), and the above condition means that all nuclear trajectories are closed. 

The topological nature of the transported charge is spelled out as follows. Suppose  that we continuosly deformate these closed trajectories: the corresponding $\zz(t)$ curve in the complex plane is accordingly deformated, but its winding number cannot vary insofar as $\zz(t) \neq 0$. The topological invariant is ``protected'' by the condition
 \[ |\zz(t)| = |\me{\Psi(t)}{\emi{\frac{2\pi}{L}\sum_{j=1}^{N^{(\rm el)}} x_j}}{\Psi(t)}| \neq 0 \label{lambda} \] at all times and in the $L \rightarrow \infty$ limit. Not surprisingly, 
  this is precisely the most general condition which characterizes the insulating state of matter;\cite{rap107,rap_a25,Stella11,rap_a33} furthermore $|\zz|=0$ even at finite $L$ for independent-electron (i.e. Kohn-Sham)  metals. See Appendix \ref{sec:insu} for more details. 
  The above findings are nothing else than a reformulation of Thouless theorem\cite{Thouless83} in a many-body framework; a somewhat different proof of the same result was provided by Niu and Thouless long ago.\cite{Niu84,Xiao10}
  
The condition that $\R_\ell(T) = \R_\ell(0)$ for all $\ell$ is a kind of ``rare event'', which would require a very long simulation time.
Nonetheless the above topological argument holds for a much larger class of trajectories. It is enough to require that the path of $\zz(t)$ in the complex plane remains a closed curve, i.e.   $\zz(T) = \zz(0)$, an event that definitely may happen in a much shorter simulation time; furthermore, since both $L$ and $T$ are macroscopic, we may replace at any time the instantaneous current $I_x(t)$ with its average $Q/T$. An additional argument---originally due to Grasselli and Baroni\cite{Grasselli19}---further validates the quantization of charge transport in the present  gedanken experiment.

The argument, exploited in Sec. \ref{sec:cond} below, is as follows. Suppose we start from a given nuclear configuration $\{\R_\ell(0)\}$, and we then follow the adiabatic time-evolution of the system along a path ending in a final configuration $\{\R_\ell(T)\}$ different from the initial one. For a macroscopic system, after a macroscopic time $T$, the accumulated angle will be \[ \gamma(T) - \gamma(0) \gg 2\pi .  \label{gb1} \] We can always ideally concatenate to the previous path another (short) path which brings back the actual configuration  $\{\R_\ell(T)\}$ into  the initial one $\{\R_\ell(0)\}$, by displacing all nuclei inside a single cell, and whose accumulated angle $\Delta \gamma$ is therefore smaller than $2\pi$. From the above arguments the total accumulated phase is \[ \gamma(T) - \gamma(0) + \Delta \gamma = 2\pi W,  \] with integer $W$ (winding number). The average current is \[ I_x = \frac{Q(T)}{T} = \frac{e}{2\pi T} [ \gamma(T) - \gamma(0)]  = \frac{e}{T}\left( W - \frac{\Delta \gamma}{2\pi} \right), \label{gb2} \] the $\Delta \gamma$ contribution vanishes il the large-$T$ (and large-$W$) limit: $Q(T)$ becomes integer.

\section{Topological oxidation numbers} \label{sec:topo}

Next we perform the following gedanken experiment. We pick only one of the nuclei, say the $\ell$-th, and we drive it round the torus (actually, to the nearby replica cell), thus crossing our ideal section once; the other $N-1$ nuclei move out of the way, in order to minimize the energy at any time. 
We assume that the system returns to the same identical configuration at the end of the cycle, and that all the other $N-1$ nuclei have not crossed the ideal section (except for two-way crossings). As shown above, if the electronic wavefunction  stays insulating at all times---as from \equ{lambda}---then the electrical charge $Q_\ell$ which has flown through the section in the process equals the winding number of $\zz(t)$  in the complex plane (times $e$). 

While there is no integer scalar charge to speak of at any given time $t$, the transport of a single nucleus along a closed path on the macroscopic torus is equivalent to transporting a time-independent scalar integer charge $Q_\ell$. This gedanken experiment has been proposed much earlier at the independent-electron level by Pendry and Hodges.\cite{Pendry84} 
As said above, nowadays a cyclic phenomenon like the one discussed here is called a Thouless pump, whose many-body generalization owes to Niu and Thouless.\cite{Niu84,Xiao10} Based on it, we have generalized the Pendry-Hodges argument to the case of interacting electrons.

The transported charge is path-independent. Consider in fact a couple of closed paths having the same initial and end points: they can be continuously deformed and made coincident without varying their winding numbers if no metallic state is crossed in the process (see also Sec. \ref{sec:pit}). The two winding numbers must therefore be equal. 

In order to identify $Q_\ell/e$ with the oxidation number $q_\ell$ of the chemical species of the $\ell$-th nucleus, one has to additionally  show that two nuclei having the same atomic number $Z$ transport the same charge $Q_\ell$ in a single cycle of the Thouless pump. To this aim, I generalize once more an argument due to Grasselli and Baroni,\cite{Grasselli19} based on additivity. If the adiabatic time $T$ is the period of the Thouless pump, then \equ{add} yields $\gamma_\ell(T) = 2\pi q_\ell$, with integer $q_\ell$ (winding number), while instead $\gamma_{\ell'}(T) = 0$ for all $\ell' \neq \ell$. Suppose next that two nuclei, labeled with $\ell=1$ and $\ell=2$, having the same atomic number $Z$, are driven along two paths: nucleus 1 from $\R_1(0)$ to $\R_1(T)$ with winding number $q_1$ in the complex $\zz$ plane. and nucleus 2 from $\R_2(0)$ to $\R_2(T)$ with winding number $q_2$. If the initial positions $\R_1(0)$ and $\R_2(0)$ can be exchanged without crossing a metallic state, then necessarily $q_1 = q_2$.

It is worth stressing that all topological definitions based on the previous Refs. \onlinecite{Pendry84,rap_a29,Jiang12,Grasselli19}, as well as the one adopted here, are not static: topological oxidation numbers are {\it dynamical} quantities, in the adiabatic limit.

\section{Independent electrons and Wannier functions} \label{sec:indep}

The case of independent electrons (in the mean-field sense) was dealt with before;\cite{Pendry84,Grasselli19} here we retrieve it as a special case of the general many-body formulation; some more details are given in Appendix \ref{sec:chern}. We consider our system and its periodic replicas as a quasi-one-dimensional ``crystal'' of macroscopic lattice constant $L$, with $N$ nuclei and $N^{\rm (el)}$ electrons per supercell. In the noninteracting (e.g. Kohn-Sham) case the insulating singlet ground state $\ket{\Psi(t)}$ can be written as a Slater determinant of $N^{(\rm el)}/2$ doubly occupied Bloch orbitals, or equivalently of $N^{(\rm el)}/2$ doubly occupied Wannier functions (WFs). In the latter case, the electronic density is partitioned into localized contributions: it is then tempting to regard the electrolyte as an assembly of closed shell entities, each of them constituted by the nucleus and by the WFs closest to it. This viewpoint would yield a definition of the oxidation numbers; nonetheless, while this picture may reasonably hold in a strongly ionic case (e.g. molten alkali halides), partitioning the electronic charge on the basis of WF proximity leads in general to difficulties and ambiguities. The issue is discussed in Ref. \onlinecite{Jiang12}, which provides a solid-state definition of oxidation numbers, based on topology ({\it not} on WF proximity), and similar in spirit to the approach of Ref. \onlinecite{Grasselli19}, generalized here.

We observe that, while the total electronic density is gauge-invariant, its partitioning into WF contributions is gauge-dependent, hence strongly nonunique. Nonetheless the contribution to macroscopic polarization from the WF centers {\it in the central cell}---as displayed next---is gauge-invariant and unique.\cite{Vanderbilt} Let us indicate as $X_j^{(\rm W)}$ the $x$-coordinates of the $N^{(\rm el)}/2$ centers. As outlined in Appendix \ref{sec:chern}
the single-point Berry phase of \equ{single} transforms (in the large-$L$ limit) into the Berry phase of the Bloch orbitals; the latter in turn can be equivalently recast in terms of WFs.\cite{Vanderbilt} The polarization assumes then a somewhat transparent expression in terms of the WF centers; when the nuclear contribution is included, the result is\cite{Vanderbilt} \bea j_x(t) &=& \dot P_x(t) = \frac{e}{2 \pi A} \dot \gamma(t) \nn &=& \frac{e}{LA} \left[\sum_{\ell=1}^N Z_\ell \dot X_\ell(t) - 2 \sum_{j=1}^{N^{\rm(el)}/2} \dot X_j^{(\rm W)}(t) \right]  . \eea This expression is similar in form to the classical case of \eqs{trivial}{gcl}, and therefore $\gamma(t)$ be recast in a similar way: \[ \gamma(t) = \mbox{Im ln }\ei{\frac{2\pi}{L} [ \,\sum_{\ell=1}^N  Z_\ell X_\ell(t) -2 \, \sum_{j=1}^{N^{\rm(el)}/2} X_j^{(\rm W)}(t) \,] } . \label{gw} \] Notice that the correspondence between nuclear coordinates and WF centers is ambiguous, and therefore 
 in general $\gamma(t)$ is not additive (only its nuclear term is such); nonetheless, as observed above, its time-derivative is additive. 
 
The same gedanken experiment described above---when specialized to the independent-electron case and formulated in terms of WFs---amounts to transporting one nucleus which ``drags'' some of the Kohn-Sham WFs, thus transporting  an even number of electrons: the oxidation number of a given chemical species obtains then from WF counting. The same occurs in the solid-state topological definition of oxidation numbers; only double-occupancy has been considered here, while in some circumstances spin-polarized ground states must  be considered as well.\cite{Jiang12}
    
\section{Ionic conductivity} \label {sec:cond}

The Green-Kubo formula for the static ionic conductivity is:\cite{Hansen} \[ \sigma = \frac{1}{L A \, k_{\rm B}T} \int_0^\infty dt\; \ev{I_x (t)\, I_x(0)} , \label{hansen} \] where the bracketed quantity is the autocorrelation function of the fluctuating charge current in absence of the electric field; the fluctuations are averaged over an equilibrium ensemble and the thermodynamic limit is understood.
 
Our previous results lead to recasting $\sigma$ as \[ \sigma  = \frac{L}{A \, k_{\rm B}T} \frac{e^2}{4\pi^2} \int_0^\infty dt\; \ev{\dot \gamma (t)\, \dot \gamma(0)} , \label{my2}  \] where $\gamma(t)$ is given in \equ{gcl} for a classical molten salt, in \equ{single} for a system of interacting electrons and classical nuclei, and in \equ{gw} in the mean-field (e.g. Kohn-Sham) approximation. In all three cases $\sigma$ is given by the autocorrelation function of the derivative of $\gamma(t)$; about the autocorrelation function of $\gamma(t)$ itself see Appendix \ref{sec:infra}.

It has been stressed above that the ab-initio current can be written in two formally equivalent ways (in the large-system limit); for the sake of clarity we rewrite the two expressions here \[ I_x(t) = \frac{e}{2 \pi} \dot\gamma(t) = \frac{e}{L} \sum_{\ell=1}^N \sum_\beta Z^*_{\ell,x\beta}(t) \dot R_{\ell,\beta}(t) . \]  Actual simulations have implemented the Berry phase expression,\cite{Cavazzoni99} the equivalent Wannier-function formulation,\cite{Rozsa18} and the   alternative expression in terms of the Born tensor, where the latter have been computed on the fly  from density-functional perturbation theory.\cite{Baroni01,Giannozzi09} The single-point Berry phase $\gamma(t)$ is implemented in a different (though similar) context: see Appendix \ref{sec:infra}.

We pause to observe that actual simulations implement the fluctuation formulas in a different geometry from the one adopted here for presentation purposes. In fact they are routinely performed in a cubic supercell of volume $L^3$  and \equ{hansen} is equivalently cast as\[ \sigma = \frac{L^3}{ 3 k_{\rm B}T} \int_0^\infty dt\; \ev{{\bf j} (t) \cdot {\bf j} (0)} , \label{cond} \] where PBCs with period $L$ are adopted on all Cartesian coordinates; the result is converged whenever $L$ is larger than both correlation lengths and diffusion lengths.  \equ{cond} is implemented in either classical molecular dynamics simulations (where the particles in the cell are ions)\cite{Frenkel,Hansen,Hansen75} or in ab-initio molecular dynamics (where the particles in the cell are electrons and nuclei, and the adiabatic approximation is adopted).\cite{French11,Grasselli19}

A surprising discovery was heuristically made in 2011 by French, Hamel, and Redmer,\cite{French11} who addressed the conductivity of partially dissociated water: upon replacing the Born tensors $Z^*_{\ell,\alpha\beta}$ with the corresponding oxidation numbers (i.e. $-2$ for oxygen and $+1$ for hydrogen) along the ab-initio trajectories, the  ab-initio $\sigma$ value was reproduced to high accuracy. Remarkably, the average of $Z^*_{\ell,\alpha\beta}$ along the trajectories was found to be a scalar differing by about 30\% from the above integer values. The reason for such counterintuitive result was found much later by Grasselli and Baroni,\cite{Grasselli19} and is rooted once more in topology. 

Here I am going to adopt the compact many-body formalism of the present work in order to present the major  result of Ref. \onlinecite{Grasselli19}, which is based on the Einstein-Helfand relationship.\cite{Helfand60} Within the geometry adopted here the Green-Kubo formula,  \eqs{hansen}{my2}, is equivalent to  the Einstein-Helfand formula \[ \sigma = \frac{L}{A k_{\rm B} T} \frac{e^2}{4 \pi^2}\, \lim_{t \rightarrow \infty} \frac{\ev{\gamma^2(t)}}{2t} . \label{eh}\] From the dynamical partitioning of \equ{add} we get \[ \gamma(t) = \gamma(0) + \sum_{\ell=1}^N \gamma_\ell(t) , \label{accu} \] where each $\gamma_\ell(t)$ is a function of the $\ell$-th trajectory. We also define the auxiliary phase angles $\tilde\gamma_\ell(t)$, where the Born effective charge tensors in \equ{dyn} are replaced by the corresponding topological oxidation numbers, over the same trajectories: \[ \tilde\gamma_\ell(t) = \frac{2\pi}{L}  q_\ell X_\ell(t) . \]The same reasoning leading to \eqs{gb1}{gb2}, concerning the total phase $\gamma(t)$, can be adopted here for the $\ell$-th phase: in the large-$t$ limit both $\gamma_\ell(t)$ and $\tilde\gamma_\ell(t)$ are unbounded, and they differ by a bounded term. Their sums differ by a bounded term as well; therefore if in \equ{accu} we replace $\gamma_\ell(t)$ with $\tilde\gamma_\ell(t)$ for all $\ell$ the limit in \equ{eh} is unaffected. 

I have provided therefore the many-body generalization of the Grasselli-Baroni theorem,\cite{Grasselli19} who also validated it by means of ab-initio molecular dynamics simulations on molten KCl; as said above, the theorem is also validated by the previous serendipitous discovery of Ref. \onlinecite{French11}.

I take the present occasion for stressing an outstanding implication of the Grasselli-Baroni theorem. Let us consider for instance liquid (undissociated) water under normal conditions: it has a finite static dielectric constant ($\epsilon_0 \simeq 80$). Hence the Faraday experiment would give a negative result, in agreement with the fact that the oxidation number of a neutral molecule as a whole is zero: it is then kind of obvious that ionic conductivity must vanish. This fact is apparently less obvious when analyzed from  the fluctuation-dissipation-theorem viewpoint. Water is strongly infrared-active,\cite{Silvestrelli97} the Born charge tensor of the molecule as a whole is quite sizable,\cite{rap116} and  the fluctuating equilibrium current is therefore conspicuously nonzero. Yet, given that the topological charge of a neutral undissociated molecule is zero, topology also guarantees that the time-integrated equilibrium autocorrelation function of the current vanishes, as indeed needed on physics considerations.

\section{More about oxidation numbers} \label{sec:more}

In inorganic chemistry the oxidation numbers (also called oxidation states) are determined via a postulatory method based on an agreed set of rules. In simple cases---when no $d$ shells are involved---the oxidation numbers are simply determined by the octet rule: atoms are assigned an octet in order of decreasing electronegativity until all valence electrons are distributed,\cite{Walsh18} except for atoms which are obviously assigned a pair (H, He, and Li). The rule is unambiguous, given that the ordering of the (simple) elements as provided by different ionicity scales is the same. Oxidation numbers are formal quantitites not related to the wavefunction or to any quantum-mechanical operator; in molecules and in crystalline solids they are not measurable, yet they correlate to several measurable properties in a postdictive way.\cite{Walsh18,rap142,rap_a29,Sit11,Jiang12} 

Given the above, it is quite remarkable that oxidation numbers in insulating liquids {\it are} instead directly measurable quantities thanks to Faraday's experiment. The fundamental difference which makes the oxidation numbers nonmeasurable in crystalline insulators and measurable in liquid insulators is that in the latter case the nuclei travel over macroscopic distances: as emphasized by Pendry and Hodges\cite{Pendry84} back in 1984, integer charges manifest themselves only when transported, owing to Thouless theorem;\cite{Thouless83,Niu84,Xiao10} static charges do not have a first-principle definition in molecules,\cite{Meister94} in solids, and in liquids.

So far, we have tacitly addressed ionic liquids where the oxidation numbers---upon chemical intuition---are quite robust in the given compound and do not depend on the fluctuating environment: this happens, for instance, when the octet rule is enough to determine the conventional oxidation numbers. There are challenging cases, though, well known in the solid state, like e.g. charge-ordering phenomena in transition metal oxides. In such cases a given element may assume different oxidation numbers depending on the ligands and on coordination. As an example, Mn can be labeled with up to seven oxidation numbers, from Mn(I) to Mn(VII), distinguished from their distinct spectroscopic and magnetic signatures.\cite{Walsh18} It is easy to guess that when the (conventionally defined) oxidation numbers do depend on the environment the liquid is not insulating and topology cannot be invoked. A challenging case of different kind is dealt with in the next Section.

The ``insulating'' qualification deserves some comments. The main phenomenological quantity which discriminates insulators from metals is $\varepsilon_\infty$, called ``high frequency'' dielectric constant:\cite{Ashcroft} it is finite in insulators and formally infinite in metals. Under the action of a macroscopic electric field, and when all nuclei are ideally ``clamped'', the electrons in a metal undergo free acceleration, while in an insulator they display  a finite polarization. Equivalently, the inverse inertia of the many-electron system is finite in metals and vanishing in insulators (see Appendix \ref{sec:insu}). 

In a liquid all intensive properties---including $\varepsilon_\infty$---are ensemble averages; the fact that $\varepsilon_\infty$ is finite does not always guarantee that all configurations visited by the system are insulating, as required by topology. Indeed, a dielectric medium with metallic inclusions  (nanodomains or mesodomains) behaves macroscopically as an insulator. One could imagine that in some circumstances metallic nanodomains may be present even in the fluctuating configurations of the insulating liquid, in which case ionic transport would violate Faraday law. 

Faraday's measurement of oxidation numbers in insulating liquids is possible owing to mass transport over macroscopic distances; the same may occur in solid-state ionic conductors. In a common
crystalline solid, instead, no mass transport may occur and oxidation numbers are not measurable in a real experiment. 

Notwithstanding, it is possible to devise a gedanken experiment sharing the same topological features as Faraday's experiment. In fact, back in 2012 Jiang et al.\cite{Jiang12} have proposed a rigorous quantum-mechanical definition of oxidation numbers in crystalline insulators based on the seminal work of Ref. \onlinecite{Pendry84} and on polarization theory: their topological approach is conceptually related to the present one.
In a crystal, when a sublattice is displaced by a lattice constant along an insulating path, the trasported charge is an integer multiple of $e$, thus defining the oxidation number of the given chemical species in the given solid.  The approach is demonstrated on a few test cases by means of density-functional calculations, where---as observed above---evaluating the oxidation numbers amounts to counting the number of transported Kohn-Sham WFs. One test case is very simple: rocksalt LiH. Not surprisingly the topological approach yields for LiH the same result as the octet rule (oxidation numbers $-1$ for H and $+1$ for Li). The other two test cases concern a given transition metal atom which assumes different oxidation numbers depending on the crystalline environment: Ba either $+3$ or $+5$, Fe either $+2$ or $+3$. In all cases explicitly dealt with, the topological theory of Ref. \onlinecite{Jiang12} yields oxidation numbers in agreement with conventional wisdom and with previous work. 
 
In Sec. \ref{sec:indep} we have assumed a singlet ground state with doubly occupied Kohn-Sham WFs; it  is then worth noticing that---in order to account for elements which may display even and odd oxidation numbers---one needs a spin-polarized ground state where some of the WFs are singly occupied, as assumed in Ref. \onlinecite{Jiang12}.
Common  quantum-chemistry wisdom\cite{SzaboOstlund} holds that spin-polarized Slater determinants may display the qualitatively correct behavior (at the price of breaking spin symmetry) in cases where double-occupancy determinants would be inadequate.  

\section{A pitfall} \label{sec:pit}

We have emphasized above the fact that charge transport in a metal occurs independently from mass transport; one may wonder whether even in some insulating liquids charge transport (over a macrocopic distance) may occur without any associated mass transport. A case study in this class has been investigated long ago: a dilute metal-alkali halide solution, i.e. a nonstoichiometric molten salt. The simulations, of an hybrid quantum-classical type, have been performed for a system of 32 K atoms and 30 Cl atoms.\cite{Fois88} The octet rule\cite{Walsh18} would assign the oxidation number $-1$ to all Cl atoms, while 30 of the K atoms would get $+1$, and two of them zero. But such conventional partitioning is grossly inadequate: snapshots from the simulation show that a singlet lone pair (dubbed bipolaron) typically sits at a Cl vacancy; the solvated electron pair moves over macroscopic distances without any associated mass transport, driven by the wandering vacancy.

A recent paper by Pegolo et al.\cite{Pegolo20} deals---among other things---with the same case study as above, by means of ab-initio molecular dynamics simulations: the aim is to assess which of the conditions warranting topological transport---and ultimately Faraday law---are actually violated. After checking that the liquid is indeed insulating, the authors perform a revealing computer experiment based on a Thouless pump. A K nucleus is adiabatically driven along two different paths beginning and ending with the same nuclear configuration; only one nucleus is driven along the loop (i.e. to the nearest cell), while the other nuclei remain in the central cell. The measured topological charge is not the same: $Q=+e$ on one path, and $Q=-e$ on the other. This means that the two paths cannot be continuously deformed and made coincident without crossing a metallic region, thus violating one of the conditions detailed above in Sec. \ref{sec:topo}. It is worth further specifying that---at the independent-electron level---a configuration is by definition metallic whenever the spectral gap closes. In the most general case we have seen from \equ{lambda} above that a configuration is metallic whenever $|\zz|$ vanishes in the large-$L$ limit (see also Appendix A).

Remarkably, the existence of metallic regions in the configuration space manifests itself even if the actual trajectories do not visit such regions. The situation is quite similar to the well known case of conical intersections in molecules, which affect the nuclear behavior even when the nuclear trajectories do not actually visit the degeneracy points (or regions): the effect goes under the name of "molecular Aharonov--Bohm effect".\cite{Mead80,Mead92}

\section{Conclusions} \label{sec:conclu}

Faraday clearly endorsed the concept of an atom as a ``small particle'' much in advance with the common wisdom of his time.\cite{Faraday} Nowadays the atomistic nature of Faraday's law is obvious; nonetheless it is too simplistic to assume that an electrolyte is a liquid where some ions are endowed with an integer charge: a condensed matter system---either solid or liquid---is an assembly of nuclei and electrons: there is no way of detecting integer-charged ions in a ``snapshot'' of the system at a given time, because the electronic density in the liquid is a continuous function. Only isolated ions in vacuo do have a static integer charge.

There is another occurrence where integer numbers appear in modern chemistry as a label for atoms: these are the oxidation numbers (also known as oxidation states). They are nonmeasurable quantities, determined according to an agreed set of rules.\cite{Walsh18} It is therefore worth stressing that Faraday's experiment is the only known instance where oxidation numbers are {\it directly} measured. This may happen because of two outstanding reasons: (i)  the charges measured by Faraday's experiment are not static, they are dynamical charges in the adiabatic regime; (ii) charges are transported over macroscopic distances, a class of phenomena where topology plays a major role.\cite{Thouless83,Niu84,Xiao10}

Over the years, only a very small number of papers was devoted to the topological interpretation of Faraday's experiment,\cite{Pendry84,rap_a29,Grasselli19} as well as to the consequent relationship between Faraday's ``equivalent numbers'' and oxidation numbers as defined in modern chemistry: this work is generalized here, by adopting a many-body formulation which additionally prompts for very compact notations. 

Besides the  time-integrated transported charge, another key observable in insulating liquids is the value of ionic conductivity at a given temperature, which obtains from the fluctuation-dissipation theorem.\cite{Hansen,Frenkel}  Here I have adopted the same logic as in Grasselli and Baroni\cite{Grasselli19} in a compact many-body framework: it is shown that when the real current is replaced with the fictitious current carried by the topological oxidation numbers (times $e$) along the real nuclear trajectories, topology warrants that the conductivity value is correctly reproduced. This interesting result was serendipitously obtained, with no explanation, in a previous ab-initio simulation on dissociated water at extreme conditions.\cite{French11}

An outstanding corollary of the Grasselli-Baroni theorem (and of its present generalization) addresses molecular undissociated liquids. Given that the topological oxidation number of a molecule as a whole is zero, the fluctuation-dissipation theorem\cite{Hansen,Frenkel} yields a vanishing  conductivity, even in presence of quite sizable fluctuating currents.

\section*{Acknowledgments}

I have discussed thoroughly the present findings with F. Grasselli, S. Baroni, and A. Marrazzo; their invaluable contribution is gratefully acknowledged. Work supported by the Office of Naval Research (USA) Grant No. N00014-20-1-2847.

\section*{Data availability}
Data sharing is not applicable to this article as no new data were created or analyzed in this study.

\appendix

\section{The insulating state of matter} \label{sec:insu}

The electronic state vector $\ket{\Psi(t)}$ addressed throughout in the present work is the instantaneous ground eigenstate of the many-body Hamiltonian \[ \hat{H}(t) = \frac{1}{2m} \sum_{i=1}^{N^{(\rm el)}} p_i^2 + \hat{V}(t) , \label{h1} \] where $\hat{V}(t)$ comprises the nuclear potential and the electron-electron interaction. In our geometry the many-body wavefunction is bounded in the $y_i,z_i$ coordinates, and obeys PBCs with period $L$ in the $x_i$ coordinates; we neglect the irrelevant $t$-dependence in the following. It is expedient, following a famous paper of Kohn's,\cite{Kohn64} to generalize \equ{h1} by including a constant vector potential (i.e. a pure gauge) along $x$: \[ \hat{H}_{\kappa} = \frac{1}{2m} \sum_{i=1}^{N^{(\rm el)}} (\, {\bf p}_i + \hbar \kappa {\bf e}_1 \,)^2 + \hat{V} , \label{h2} \] where ${\bf e}_1$ is the unit vector in the $x$ direction, and $\kappa$ is a scalar with the dimensions of an inverse length. Let $\ket{\Psi_\kappa}$ be the ground PBC eigenstate of $\hat{H_\kappa}$ with energy $E_\kappa$. If we define $\hat{x} = \sum_i x_i$, the quantity $\emi{\kappa \hat{x}} \ket{\Psi_0}$ is a solution of Schr\"odinger equation with energy $E_0$, but it does not obeys PBCs (for a generic $\kappa$): hence it is not an eigenstate. The genuine PBCs ground eigenstate $\ket{\Psi_\kappa}$ has a nontrivial $\kappa$ dependence.

The kinetic-energy term in \equ{h2} yields the many-body velocity operator (along $x$): \[ \hat v_\kappa = \frac{1}{m} \sum_{i=1}^{N^{(\rm el)}} ({p}_{ix} + \hbar \kappa ) = \frac{1}{\hbar} \frac{d}{d\kappa} \hat{H}_\kappa , \] and the stationary electronic current which crosses a vessel's section is \[ I_\kappa^{(\rm el)} = - \frac{e}{\hbar L} \me{\Psi_\kappa}{ \frac{d}{d\kappa} \hat{H}_\kappa }{\Psi_\kappa} =  - \frac{e}{\hbar L}  \frac{d E_\kappa }{d\kappa} . \] Given that no stationary current may flow in an insulator when the nuclei are at rest, the ground state energy is $\kappa$-independent. Matters are different in metals: it has been shown by Kohn\cite{Kohn64} that $d^2 E_\kappa/d \kappa^2$ is proportional to the Drude weight. We remind that the Drude weight measures the inverse inertia of the many-electron system in response to a constant electric field; in absence of dissipation, the electrons in a metal undergo free acceleration.\cite{rap157}

Whenever $\kappa$ is an integer multiple of $2\pi/L$ the state vector $\emi{\kappa \hat{x}} \ket{\Psi_0}$ obeys PBCs, besides being a solution of Schr\"odinger equation with energy $E_0$: it is therefore the ground eigenstate of $\hat{H}_\kappa$. We pick one of these values: $\kappa_1=2\pi/L$.
According to Kohn's prescription, the state vector $\ket{\Psi_{\kappa_1}}$ obtains by adiabatically following $\ket{\Psi_0}$ when $\kappa$ is adiabatically turned on. Because of what has been said above, $\ket{\Psi_{\kappa_1}}$ is orthogonal to $\emi{\kappa_1 \hat{x}} \ket{\Psi_0}$ in metals, while it coincides with it in insulators (apart for an arbitrary phase factor), ergo: \bea |\me{\Psi_{\kappa_1}}{\emi{\frac{2\pi}{L}\sum_{j=1}^{N^{(\rm el)}} x_j}}{\Psi_0}| &=& 0 \qquad \mbox{in metals}; \\ |\me{\Psi_{\kappa_1}}{\emi{\frac{2\pi}{L}\sum_{j=1}^{N^{(\rm el)}} x_j}}{\Psi_0}| &=& 1 \qquad \mbox{in insulators} . \eea 
By approximating $\ket{\Psi_{\kappa_1}}$ with $\ket{\Psi_{0}}$ one gets $|\zz|$ as defined in \equ{lambda}: \[ | \me{\Psi_{\kappa_1}}{\emi{\frac{2\pi}{L}\sum_{j=1}^{N^{(\rm el)}} x_j}}{\Psi_0}| 
\simeq | \me{\Psi_0}{\emi{\frac{2\pi}{L}\sum_{j=1}^{N^{(\rm el)}} x_j}}{\Psi_0}| = |\zz| . \] The approximation is to leading order in $1/N^{(\rm el)}$, i.e. to leading order in $1/L$; in the large system limit $|\zz|$ tends to one from below in all insulators, and to zero in all metals, as stated in the main text. It is worth noticing that in the special case of noninteracting electrons (either Hartree-Fock or Kohn-Sham) a selection rule guarantees that for metals $\zz = 0$ even {\it at finite size};\cite{rap107,rap_a33} the large-size limit is instead needed for interacting electrons.\cite{Stella11}

\section{Infrared spectra} \label{sec:infra}

This paper is devoted to two static properties (namely, charge transport and dc ionic conductivity), emphasizing their topological nature. For the sake of completeness, I discuss here an important $\omega$-dependent property which is closely related to them, and is instead merely geometrical.

Infrared spectra are expressed in terms of the imaginary part of the dielectric function $\varepsilon(\omega)$. In order to show the close relationship to the previous results, I adopt the same cylindrical geometry and the same notations as above, in which case the known expression\cite{Guillot91} can be recast in the form \bea \mbox{Im } \varepsilon(\omega) &=& \frac{2 \pi LA \omega}{k_{\rm B} T}\int_{-\infty}^\infty dt \; \ei{\omega t} \ev{P_x(t)\,  P_x(0)} \nn &=& \frac{e^2 L \omega}{2 \pi A  k_{\rm B} T} \int_{-\infty}^\infty dt \; \ei{\omega t}  \ev{\gamma(t) \gamma(0) }; \label{infra} \eea as above, the linear dimensions of the system must be  larger than both correlation lengths and diffusion lengths. The main entry in \equ{infra} is the power spectrum of the autocorrelation function of $\gamma(t)$ itself, while we remind that instead the main entry in the expression of ionic conductivity is the autocorrelation function of the {\it time derivative} of $\gamma(t)$, as in \equ{my2}.

The single-point Berry phase $\gamma(t)$, as defined in \equ{single}, is actually implemented since long time in ab-initio simulations in order to compute the infrared spectra of liquids and amorphous solids from \equ{infra}.\cite{Silvestrelli97,Debernardi97}  
Equivalently the WF expression, \equ{gw}, has been adopted in the literature in the paradigmatic case of liquid water,\cite{rap129,rap140} where hydrogen bonding is at the root of unusual dielectric properties.\cite{rap134}

\section{Winding number as a Chern number} \label{sec:chern}

As said in Sec. \ref{sec:indep}, in the independent-electron case (Hartree-Fock or Kohn-Sham) our macroscopic vessel is regarded as a quasi-one-dimensional ``crystal'' of macroscopic constant $L$, with $N$ nuclei and $N^{\rm (el)}$ electrons per supercell. The insulating ground state can be written as a Slater determinant of doubly occupied Bloch orbitals $\ket{\psi_{jk}}$, with a gapped spectrum and $N^{(\rm el)}/2$ filled bands; the quasi-one-dimensional orbitals \[ \ket{u_{jk}} = \emi{kx} \ket{\psi_{jk}}, \quad k \in [0,2\pi/L), \quad j=1,\dots N^{(\rm el)}  \] are periodic in $x$ and bounded in $(y,z)$.

It is shown elsewhere\cite{rap100,rap_a20,rap162} that the single-point Berry-phase
\[ \gamma^{(\rm el)}(t) = \mbox{Im ln } \me{\Psi(t)}{\emi{\frac{2\pi}{L} \sum_{j=1}^{N^{(\rm el)}} x_j}}{\Psi(t)} \label{s2} \] yields---in the crystalline case---the standard Berry phase of polarization theory:\cite{Vanderbilt} \[ \gamma^{(\rm el)}(t) = 2 \; i \int_0^{2\pi/L} dk \sum_{j=1}^{N^{(\rm el)}/2} \ev{u_{jk}(t) | \frac{\partial}{\partial k}  u_{jk}(t)} ,\] where $\ket{\psi_{jk}}$ is chosen within the periodic gauge, ergo \[ \ket{u_{j,k+G}} = \emi{Gx}\ket{u_{jk}}, \qquad G=2 \pi/L\label{periok} .\] It is expedient to define the (dimensionless) differential form \[ {\cal A} = i\!\!\sum_{j=1}^{N^{(\rm el)}/2}\!\! [\ev{u_{jk}(t) | \frac{\partial}{\partial k}  u_{jk}(t)} dk +  \ev{u_{jk}(t) | \frac{\partial}{\partial t}  u_{jk}(t)} dt  ], \label{A} \] known in mathematical speak as a Chern-Simons 1-form, and its exterior derivative $\Omega=d \wedge {\cal A}$  (first Chern form) \bea \Omega &=&  \Omega_k(t)\;  d k \, dt , \nn \Omega_k(t) &=&- 2 \; \mbox{Im} \sum_{j=1}^{N^{(\rm el)}/2}\ev{\frac{\partial}{\partial k} u_{jk}(t) | \frac{\partial}{\partial t}  u_{jk}(t)} , \label{Omega} \eea  where $\Omega_k(t)$ is a Berry curvature in the $(k,t)$ variables.

\begin{figure}[t]
\centering
\includegraphics[width=0.3\linewidth]{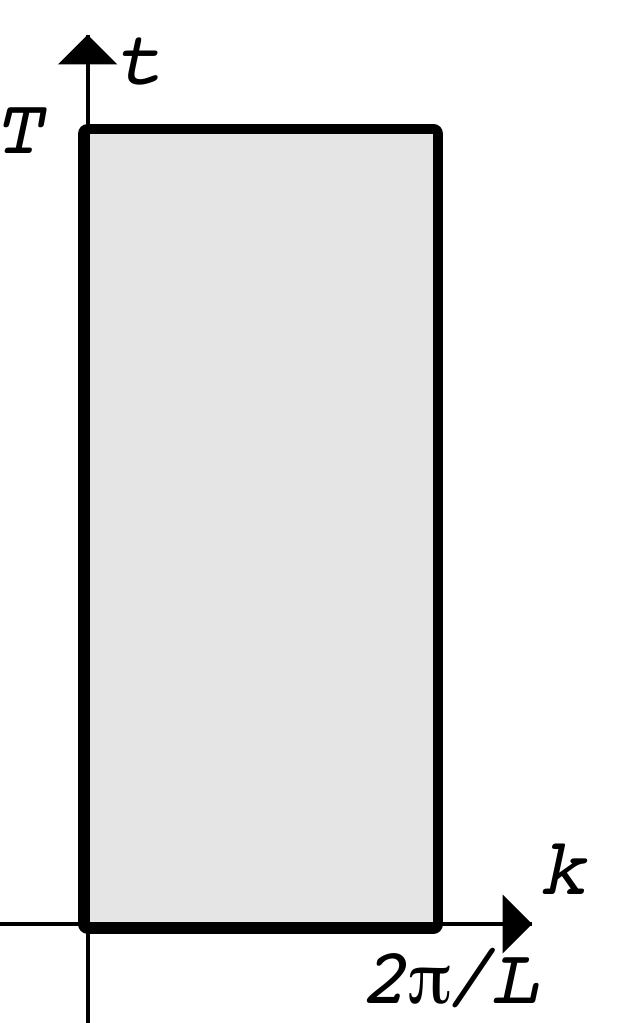} 
\caption{The many-band Chern number $C_1$ is the integral of \equ{Omega} over the rectangular domain in the $(k,t)$ plane, divided by $2\pi$. Its value is $L$-independent, and in the $L \rightarrow \infty$ limit it coincides with the winding number $W$ (electronic term thereof) thoroughly addressed in this work.}
\label{fig} \end{figure}

As shown in the main text, whenever the Hamiltonian is periodic in $t$ (with period $T$), then \[ W^{(\rm el)} = \frac{1}{2\pi} [\gamma^{(\rm el)}(T) - \gamma^{(\rm el)}(0) ]  \label{W}\] is integer (many-electron winding number). 
With reference to Fig. \ref{fig}, it is clear that $2 \pi W^{(\rm el)}$ is equal to twice the integral of ${\cal A}$, \equ{A}, on the top horizontal segment, minus its analogue over the bottom segment. Furthermore, the two integrals of ${\cal A}$ over the two vertical segments cancel owing to \equ{periok}, and we thus get \[ \frac{1}{2}W^{(\rm el)} = - \frac{1}{2\pi} \oint {\cal A}, \] where the (counterclockwise) loop is the contour of the rectangle in the $(k,t)$ plane. Owing to Stokes theorem then \[  \frac{1}{2}W^{(\rm el)} = - \frac{1}{2\pi} \int_0^{2\pi/L} dk \int_0^T dt \;  \Omega_k(t) = C_1 , \label{Thouless} \] where $C_1 \in {\mathbb Z}$ is the first Chern number; in mathematical speak the integration domain is a compact orientable manifold (a torus, owing the the periodicity in both $k$ and $t$).

The second equality in \equ{Thouless} is a statement of the celebrated Thouless theorem;\cite{Thouless83,Vanderbilt} here $C_1$ is the sum of the individual Chern numbers of the $N^{(\rm el)}/2$ occupied bands. Since $N^{(\rm el)}$ is proportional to $L$, the $C_1$ value is $L$-independent, i.e. it is independent of the area of the integration domain in Fig. \ref{fig}. Notice also that in the above formulas $k$ is a continuous variable, ergo the large-system limit is implicit even at {\it finite} L; instead the single-point Berry phase acquires its exact physical meaning only in the 
$L \rightarrow \infty$ limit. One might therefore say that the electronic term $W^{(\rm el)}$ in the winding number is nothing else than (twice) the many-band Thouless' Chern number $C_1$ in the limit where the two vertical segments in Fig. \ref{fig} contract to a single segment.

The present results are worth a further interpretation. We notice that the large-$L$ limit of \equ{Thouless} yields \[ W^{(\rm el)} \longrightarrow - \frac{2}{L} \int_0^T dt \;  \Omega_0(t) ; \] comparing to \equ{W}, the (extensive) Berry curvature $\Omega_0(t)$ is related to the single-point Berry phase  as \[2 \, \Omega_0(t) = \frac{L}{2\pi} \dot \gamma^{(\rm el)}(t) . \]

%\bibliography{$HOME/inputs/huge_bib,add_bib}
%\bibliographystyle{unsrt}

\end{document}